\documentclass[12pt,a4paper]{article}
\newenvironment{figlist}{\begin{list}{\bf Figure \arabic{fig}:}
{\usecounter{fig}\setlength{\labelwidth}{0.5cm}
\setlength{\labelsep}{0.2cm}\setlength{\parsep}
{1ex plus0.2ex minus0.1ex}\setlength{\itemsep}{1ex plus0.2ex}}}
{\end{list}}
\newcounter{fig}

\begin{document}

\title{ Experimental Indication of Structural Heterogeneities 
in Fragile Hydrogen-bonded Liquids }

\author{
D. MORINEAU, C. ALBA-SIMIONESCO\\
{\small\em  CPMA URA 1104 CNRS, b\^{a}timent 490, Universit\'{e} Paris-Sud, 91405 Orsay,
France}\\ 
M.-C. BELLISSENT-FUNEL\\
{\small\em Laboratoire L\'eon Brillouin, C.E.N. Saclay 91 Gif sur Yvette,
France}\\
M.-F. LAUTHI\'E\\
{\small\em  L.A.S.I.R., 94 Thiais,
France}\\}

\maketitle

\begin{abstract}
 We present the first experimental characterization in molecular 
fragile glassformers of a "prepeak" that appears significantly 
below the main peak of the static structure factor. The temperature
and density dependences of this prepeak are studied via elastic neutron 
scattering experiments under high pressure (up to 300 MPa) in 
m-toluidine and m-fluoroaniline. The prepeak intensity 
increases with decreasing temperature, but it remains constant 
with increasing pressure while its position and width stay
roughly the same. These features are opposite to those observed for 
the "first sharp diffraction peak" of network
glasses. The origin of the phenomenon is analyzed with the 
help of Monte Carlo simulations. We associate the prepeak 
with hydrogen-bond-induced heterogeneities
(or clusters) whose limited size results from an exclusion
 effect between benzene rings that prevents the extension of
 a hydrogen-bond network. Implications for the dynamics of 
the liquid close to the glass transition are finally considered.


\end{abstract}
\clearpage

Analyzing the structure of glasses and supercooled liquids can give some 
insight into the mechanisms involved in the glass transition at a 
micro-or mesoscopic scale. Recently there has been  growing experimental 
evidence of the existence of heterogeneities that develop with 
decreasing temperature in fragile supercooled liquids \cite{ino}. 
Such heterogeneities are at the basis of many theories \cite{theo} 
of the dramatic slowing down of the relaxations when approaching 
the glass transition. Heterogeneities of some sort have also 
been invoked to explain the apparently anomalous relaxational 
properties of monohydric alcohols and supercooled water : 
these latter would be due to a clustering phenomenon involving 
hydrogen-bonding interactions \cite{caa}.
However, the evidence for heterogeneities, domains or clusters is 
mostly indirect, i.e. extracted from dynamical measurements. 
It would thus be of major interest to find the signature of 
such entities directly in structural data.

Network-forming glasses and melts show a first sharp diffraction 
peak (FSDP) in X-ray and neutron diffraction data which, 
although still debated, seems convincingly attributed to some 
intermediate-range order \cite{ell}. As for molecular systems, 
few results about the structure of their supercooled liquid and glassy 
states have been reported, and the typical neutron or X-ray 
pattern consists of broad peaks that evolve continuously with 
temperature and display no drastic changes at the glass 
temperature $T_{g}$. A recent low-angle neutron diffraction 
study has concluded to the absence of heterogeneities in 
deuterated propylene glycol \cite{nag}.

In this letter we present the first experimental characterization 
for molecular fragile glassformers of the temperature
and density dependence of a "prepeak" which appears at a wave vector
 $Q_{pp}\approx$0.5 \AA$^{-1}$, i.e., significantly below the main 
peak of the static structure factor $S(Q)$. We focus on a 
series of substituted aromatic liquids, toluene (TOL), 
m-toluidine (MTOL) and m-fluoroaniline (MFA). An interesting property
of these liquids is that crystallization can be readily 
avoided so that they are good candidates for a thorough study 
of the metastable liquid state. Despite the differences in 
their dipole moment and intermolecular H-bonds strength, these
liquids present the same thermodynamic signature at $T_{g}$ 
and are considered as some of the most fragile systems \cite{cas}.

The experiments were carried out on the 7C2 and G6.1 spectrometers at 
the reactor Orph\'ee of the laboratoire L\'eon 
Brillouin\footnote{laboratoire commun CEA-CNRS}\ (Saclay, France) at
three different incident wave lengths in order to cover a 
large $Q$-range from 0.1 
to 10 \AA$^{-1}$. An isotopic purity of the samples of more than 99\%
was obtained by catalytic H-D exchange for the fully deuterated 
 MTOL (C$_{6}$D$_{4}$CD$_{3}$ND$_{2}$), 
the partially deuterated MTOL (C$_{6}$D$_{4}$CD$_{3}$NH$_{2}$), 
and MFA (C$_{6}$HD$_{3}$FND$_{2}$). The temperature range was 
from 320 K to 77 K, from well above the melting temperature 
$T_{m}$ down to well below $T_{g}$ , and the pressure
range was from normal pressure up to 300 MPa. The pressure 
was measured out of the cell and in the sample at the bottom 
of the cell, thus allowing to check that the system is 
hydrostatically compressed. Spectra were corrected for detector 
efficiencies, background, cell contribution and multiple scattering 
using standard procedures. After normalization, Placzek corrections 
were evaluated with a polynomial function. Other details on the 
experimental setup and data reduction will be given in 
a forthcoming publication \cite{mor}.

Fully corrected static structure factors $S(Q)$ are presented in figs. 1 
and 2 for MFA and partially deuterated MTOL, respectively. Like
many aromatic liquids, a main diffraction peak is observed in the Q-range 
between 1.3 \AA$^{-1}$ and 2.0 \AA$^{-1}$ and present differences 
in shape that are mainly due to different chemical constitution 
and cancellation effects in the linear combination of the 
partial structure factors.

The most noticeable feature in the $S(Q)$ is a "prepeak" that
appears at a much lower $Q$ 
than the main peak for MFA and MTOL (figs. 1 and 2). Its intensity 
increases when $T$ decreases down to $T_{g}$, at which point 
it stays roughly constant (cf. for instance the inset in
fig. 2). As illustrated in fig. 2, the position and the width 
of the prepeak remain constant to a good approximation, 
whereas the main peak shifts slightly to higher $Q$ due
to the increase in density. By using the pressure as an additional
external parameter, one can disentangle the effects of density 
and temperature. We have thus measured the isothermal
and isochoric evolutions of the $S(Q)$ of MFA. The results 
for the $T$=285 K isotherm (liquid state) are displayed in the 
inset of fig. 1. For a density increase of about 10$\%$,
the main peak moves to higher $Q$, as expected, whereas
the prepeak appears remarkably constant in position, width, 
and intensity. The same features are also observed for 
the $T$=150 K isotherm in the glassy state.
A quite different behavior is obtained in the evolution of
$S(Q)$ with temperature at a constant density of
1.3 gr/cm$^{3}$ : as shown in fig. 1, the main peak remains
roughly constant, which indicates no significant modification 
of the short-range order and local molecular packing, but 
the intensity of the prepeak increases
by a factor of two when the temperature is decreased by 90 K. 
The fact that the prepeak in MFA and MTOL seems to be only 
temperature dependent, together with the absence
of prepeak in the $S(Q)$ of liquid and glassy TOL  at 73 K \cite{mor},
suggests that the prepeak is a consequence of hydrogen bonding 
interactions.

"Prepeaks" (or FSDP) are commonly observed in the $S(Q)$'s 
of network glasses and are usually interpreted as the signature 
of a structure in real space that has a typical length greater 
than the first shell of neighbors \cite{ell}. Recent computer simulation
studies support the idea that such an "intermediate-range order" 
originates from angular constraints in the topology of the random 
network formed by tetrahedrally coordinated atoms \cite{mad}. 
Despite the apparent similarity in the diffraction
pattern, this explanation cannot be transposed to MFA and MTOL. 
As already stressed, these latter liquids are fragile. 
The H-bonding interactions are not dominant enough in their case 
to generate a continuous network nor to change the extreme fragility
that is typical of the whole series of substituted aromatic 
liquids (as a fragility indicator, note that the parameter D of
the Vogel-Tammann-Fulcher equation is between 3 and 6). Quite 
the contrary, network-forming systems are strong (e.g., SiO$_{2}$) 
or intermediate  (e.g. ZnCl$_{2}$) glassformers. 
Several characteristic features of the prepeak are also very different
in network glasses and in the present case. In MFA and MTOL the 
position of the prepeak is at lower $Q$ than it is in the 
network systems ($Q_{pp}$=0.5 \AA$^{-1}$ compared 
to 1-2 \AA$^{-1}$).
The pressure dependences, as well as the temperature dependences, 
are also at variance since unlike in the vitreous
SiO$_{2}$ \cite{sug} the prepeaks in MFA and MTOL are 
mainly dominated by temperature and not by density changes. 
Moreover, while the main peak is sensitive 
to density changes and shifts to higher $Q$ by compressing or 
cooling, the prepeak position and its full width at half 
maximum are constant over a wide temperature
and density range, from the stable liquid down to the glassy state.
The inset of fig. 2 shows that the temperature dependence 
of the prepeak intensity in the supercooled liquid is  more pronounced
than the normal Debye-Waller dependence shown in the glassy state, 
and this is independent of the thermal treatment. Similar 
conclusions have been drawn from previous X-ray experiments 
at normal pressure \cite{desc}. 

Due to the low value of $Q_{pp}$ (one third of that of the main peak) 
and to the peculiar T and P dependences, the existence of the 
prepeak indicates a spatial organization of the molecules
that goes beyond the usual short-range liquid order.
However, its origin cannot be explained by 
pseudo-crystalline models \cite{gas} for several reasons.
Firstly, the prepeak appears in a reversible way
at temperatures above the melting point and persists in the 
stable liquid state. Secondly, for some systems a Bragg peak 
can be found at 0.5 \AA$^{-1}$ in the $S(Q)$ of the crystal
but no prepeak is observed in the $S(Q)$ of the glassy state
as it was found for TOL \cite{mor}.

In disordered systems, one often argues that the position 
of a peak reflects some repetitive characteristic distance 
between structural units and that its width corresponds
to a correlation length \cite{ber, sal}. In order to check 
the possibility of an increasing structural correlation length 
in the liquid, the $S(Q)$'s of partially deuterated MTOL obtained
at different temperatures and normal pressure have been normalized
at the prepeak position (cf. fig. 2). It clearly appears 
that the half-width at half-maximum is independent of 
temperature ; it yields a typical length of 50 \AA, but no 
increase is detected from above the melting temperature down 
to below $T_g$, i.e., over more than 200 K.

We suggest an explanation for the origin of the prepeak 
that involves the existence of H-bond-induced heterogeneities or
clusters with no significant growth of the typical length. 
In order to gain more insight, we have also carried out 
Monte-Carlo simulations of MTOL by using an effective OPLS 
potential developed by Jorgensen \cite{jorg}. Monte-Carlo 
simulations have been performed in the isothermal-isobaric 
ensemble ($N,P,T$) at different temperatures and pressures 
with a Metropolis sampling. The system consists of 768
molecules (more than 10$^{4}$ sites) in a cubic cell ca 50 \AA{} 
on a side with periodic boundary conditions.
Full details on the simulations will be provided in a 
forthcoming publication. After an equilibration period  
covering 10$^{6}$ to 3.10$^{6}$ configurations, the structural
properties were averaged over the next 2.10$^{6}$ to 
4.10$^{6}$ configurations. The total and some partial $S(Q)$'s 
and radial distribution functions (RDF) at atmospheric pressure
and room temperature are presented in figs. 3a and 3b. We have 
checked that the agreement between experiment and simulation 
is very good for the total $S(Q)$ \cite{mor}. The numerical simulations 
reveal the existence of a short-to-medium range order that shows up
in the partial structure factors associated with
 the NH$_{2}$ group, $S_{\mathrm{NN}}(Q)$ or $S_{\mathrm{NH}}(Q)$: 
the peak in $S_{\mathrm{NN}}(Q)$ or $S_{\mathrm{NH}}(Q)$ has 
a position at 0.5 \AA$^{-1}$, which  corresponds to that of 
the prepeak in the total $S(Q)$; its behavior is the same
as that observed in the experiments described above when temperature
or density is independently changed \cite{mor}, while an
isothermal compression tends to bring the peripheric groups 
(C, H, CH$_{3}$) closer and affects weakly the structure factor
of the centers of the molecules, 
$S_{\mathrm{CC}}(Q)$ \cite{dosseh}.
When one replaces the NH$_{2}$ and CH$_{3}$ groups by
H atoms in the configurations of fig. 3a, the structure
of benzene is restored \cite{chie}: the packing of the benzene 
rings is the dominant effect at  $Q$'s above 1.0 \AA$^{-1}$ 
and H-bonding effect is only illustrated at lower $Q$'s.
The two contributions are well separated in the particular 
case of MTOL and MFA giving rise to a prepeak in the total 
structure factor. Studying the center of the molecule and the 
aminogroup RDF's provides two levels of structural organization
as presented in fig. 3b. The former RDF presents   
damped oscillations with a period of 5-6 \AA, about one 
molecular diameter, and is very similar to what is observed 
in simple Van der Waals liquids \cite{dosseh}. In the latter appear first 
the distance N-N of 3 \AA, which is typical of the H-bond 
itself, and an oscillation  at 12 \AA{} (and seemingly 
at 25 \AA{} too). This oscillation occurs with a typical 
period, which is  the mean intercluster separation 
or the mean cluster diameter. By Fourier transform, this typical period
gives rise in Q-space to the prepeak at 0.5 \AA$^{-1}$.
On the basis of geometric criteria for the H-bond, we have checked 
the number of molecules connected by H-bonds in a cluster. We have found 
a size distribution of these clusters between 2 to 10 
molecules, whose averaged number increases when the 
temperature decreases while the largest size stays at 10 molecules.  
The steric exclusion between aromatic rings prevents 
the extension of a continuous H-bond network 
contrary to what occurs in intermediate H-bonded liquids
such as small alcohol molecules \cite{nag,ber}.
These clusters can be refered as structural heterogeneities
specific to the systems under study ; they do not occur in TOL since 
there is no H-bond interaction, nor in propylene glycol \cite{nag} or 
ethanol \cite{ber} where the packing constraint
is reduced. These clusters do not seemingly order 
themselves, which is supported by the experimental observation 
that the prepeak width remains essentially constant.
Because of this, the fragile character of these liquids 
is not altered and remains similar to that of the van der 
Waals TOL or the m-Xylene where no prepeaks are observed. 

The clusters observed here have little effect on the overall 
slowing down of the relaxations with temperature, since 
it is very similar for the whole series of 
substituted aromatic liquids, irrespective of the presence of
a prepeak in the $S(Q)$. Therefore, they are presumably not 
related to the heterogeneities or domains recently discussed in the 
litterature \cite{ino,theo}.

However, the presence of clustering phenomena might induce significant 
differences in the relaxation times associated with different probes.
In the case of MFA, the response under a mechanical stress is faster 
(by a factor of 3) than that under an electrical stress \cite{cut-leb}. 
In both cases, the relaxations have roughly the same $T$-dependence,
and the relaxation functions
are highly non-exponential with a Kolraush (stretching) exponent $\beta$ 
of 0.37 for shear or volume relaxations and 0.61 for dielectric 
relaxation \cite{cut-leb} at the same temperature $T$=210 K.
The non-exponential character of the relaxations
which is at variance to what is observed for monohydric alcohols 
\cite{caa} is probably related to the absence of a 3-dimensional 
network and to the polydispersity and the limited size 
of the clusters.  As suggested in ref. 3, the experimental observations 
could be rationalyzed by considering that the dielectric process 
is dominated by the overall motion of the clusters contrary 
to the mechanical relaxation. An other interpretation could arise
from recent experiments \cite{fischer} which suggest
that the slowest Debye-type process observed by dielectric 
spectroscopy for 1-propanol does is not assigned
to any structural relaxation but to distinct H-bonding effects.  
These special features due to 
the H-bonding are obviously not expected in a van der Waals 
isostructural system like m-xylene or TOL. In the latter case, 
as indeed suggested by recent work \cite{boh}, the dielectric 
relaxation mode would not be the slowest one. More work in 
this direction would be usefull to clarify the dynamical signature
of the H-bond-induced clusters in fragile liquids.
\section*{}

The authors are very grateful to 
Dr G. Tarjus  and Prof. M. Descamps for very helpful and
stimulating discussions and to Drs A. Boutin and 
R. Pellenq for their help in numerical simulations.  
The authors would like to thank R. Millet from 
L.L.B. for his constant technical assistance and support during all
the experiments and Mrs Ratovelomanana for the sample preparation. 
This work has benefit from the use of the neutron source 
of the reactor Orph\'{e}e (LLB, CEA-CNRS, Saclay) and C94 and 
C98 Cray computers facilities from IDRIS (Orsay).

\clearpage

\clearpage
\begin{center}
Figure captions \\~
\end{center}

\begin{figlist}
\item  Total static structure factor $S(Q)$ of MFA ($T_g$=170
K, not known $T_m$) along the isochore  
$\rho$=1.3 gr/cm$^3$, at $T$=195 K, 0.1 MPa (circles) and 
$T$=285 K, 260 MPa (squares). The difference between the 2 spectra 
(triangles) is only sensitive to the prepeak intensity. The inset shows
the same fonction along the isotherm T=285 K at two pressures 0.1 MPa 
(circles) and 260 MPa (squares) corresponding to a density change 
of 10\%. The difference between the 2 spectra (triangles) only shows 
a shift of the main peak position.  

\item Reduced structure factor $\Delta S(Q,T)/ \Delta S(Q=0.5,T)$ 
of the partially deuterated liquid MTOL (C$_{6}$D$_{4}$CD$_{3}$NH$_{2}$)
at 0.1 MPa for different T's corresponding to the points in the inset.
It is obtained by subtracting $S(Q, T$=310 K) from
$S(Q,T)$ and normalizing at the 
prepeak position $Q_{pp}=0.5$ \AA$^{-1}$. Note that the width 
and the position of the prepeak are constant, whereas the 
main peak changes with $T$. The inset shows $\Delta S(Q=0.5,T)$ 
as a function of $T$.

\item  Results of Monte-Carlo simulation of MTOL at $T$=291 K 
and 0.1 MPa. a) Total structure factor $S(Q)$ (circles), 
partial structure factor $S_{\mathrm{NN}}(Q)$ (full lines)
and center-to-center structure factor $S_{\mathrm{cc}}(Q)$ 
(dotted lines). Also displayed (squares) the $S(Q)$ 
for a pseudo-benzene obtained by replacing all NH$_2$ and CH$_3$ 
groups by H atoms in the configurations of MTOL. 
The low $Q$ residual shoulder appearing in the partial $S(Q)$ 
and for the pseudo-benzene is due to the Fourier transformation 
procedure.
b) N-N partial RDF (squares) and center-to-center partial RDF (rings)
presenting oscillations with two different associated periods.

\end{figlist}


\begin{thebibliography}{99}
\bibitem{ino} CICERONE M. T. {\em et al.,} {\em J. Chem. Phys.,} {\bf 104} (1996) 7210; SCHMIDT-ROHR K. and SPIESS H.W., {\em Phys. Rev. Lett.} {\bf 66} (1991) 3020; BOHMER R. {\em et al.,} 
{\em Europhys. Lett.,} {\bf 36} (1996) 55; SCHIENER B. {\em et al.,} {\em Science,} {\bf 274} (1996) 752.
\bibitem{theo} KIRKPATRICK T. R. {\em et al.,} {\em Phys. Rev. A,} {\bf 40} (1989) 1045; MEL'CUK
A. I., {\em Phys. Rev. Lett.,} {\bf 75} (1995) 2522; KIVELSON D. {\em et al.,} {\em PHYSICA A} {\bf 219} (1995) 27;
KIVELSON D. and TARJUS G., {\em Prog. Theor. Phys.} in press (1998).
\bibitem{caa} FLORIANO M. A. and ANGELL C. A., {\em J. Chem. Phys.,} {\bf 91} (1989) 2537; 
ANGELL C.A., in "Correlations and Connectivity", H. E. STANLEY, N. OSTROWSKY (eds.),
Kluwer Acad. Publish. (1990). 
\bibitem{ell} ELLIOTT S. R., {\em Nature,} {\bf 354} (1991) 445; {\em J. Phys. : Condens. Matter,}
 {\bf 4} (1992) 7661.
\bibitem{nag} LEHENY R. {\em et al.,} {\em J. Chem. Phys.,} {\bf 105} (1996) 7783.
\bibitem{cas} ANGELL C. A., p.3 in "Relaxation in Complex Systems", 
NGAI K. L. and WRIGHT G. B. (eds.) Office of Naval Research, Washington, (1984);
ALBA-SIMIONESCO C., FAN J. and ANGELL C. A., submitted to {\em J. Chem. Phys.,} (1998).
\bibitem{mor} MORINEAU D. and ALBA-SIMIONESCO C., to be published.
\bibitem{mad} WILSON M. and MADDEN P., {\em J. Phys. : Condens. Matter,} {\bf 6} (1994) A151.
\bibitem{sug} SUGAI S. and ONODERA A., {\em Phys. Rev. Lett.,} {\bf 77} (1996) 4210.
\bibitem{desc} DESCAMPS M. {\em et al.,} {\em Prog. Theor. Phys.,} {\bf 126} (1997) 207.
\bibitem{gas} GASKELL P. H., {\em Materials Science and Technology,} CAHN R. W. {\em et al.,}(ed.), {\bf 9} (1991) 175.
\bibitem{ber} SRINIVASAN A. {\em et al.,} {\em Phys. Rev. B,} {\bf53} (1996) 8172; FAYOS R. {\em et al.,} {\em Phys. Rev. Lett.,}
{\bf77} (1996) 3823.
\bibitem{sal} SALMON P., {\em Proc. R. Soc. Lond. A,} {\bf445} (1994) 351.
\bibitem{jorg} JORGENSEN W. {\em et al.,} {\em J. of Comput. Chem.,} {\bf 14} (1993) 206.
\bibitem{dosseh} MORINEAU D. {\em et al.,} {\em Molec. Sim.,} {\bf20} (1997) 95.
\bibitem{chie} BARTSCH E. {\em et al.,} {\em Ber. Bunsenges. Phys. Chem.,} {\bf 89} (1985) 147; 
Chem. Phys., {\bf 169} 373 (1993).
\bibitem{cut-leb} CUTRONI M. {\em et al.,} {\em J. Phys.: Condens.
Matter,} {\bf 6} 5283 (1994); ALBA-SIMIONESCO C., unpublished results. 
\bibitem{fischer} HANSEN C.{\em et al.,} {\em J. Chem. Phys.,}
 {\bf 107} (1997) 1086. 
\bibitem{boh} DOSS A. {\em et al.,} {\em J. Chem. Phys.,} {\bf 107}
 (1997) 1740.

\end{thebibliography}
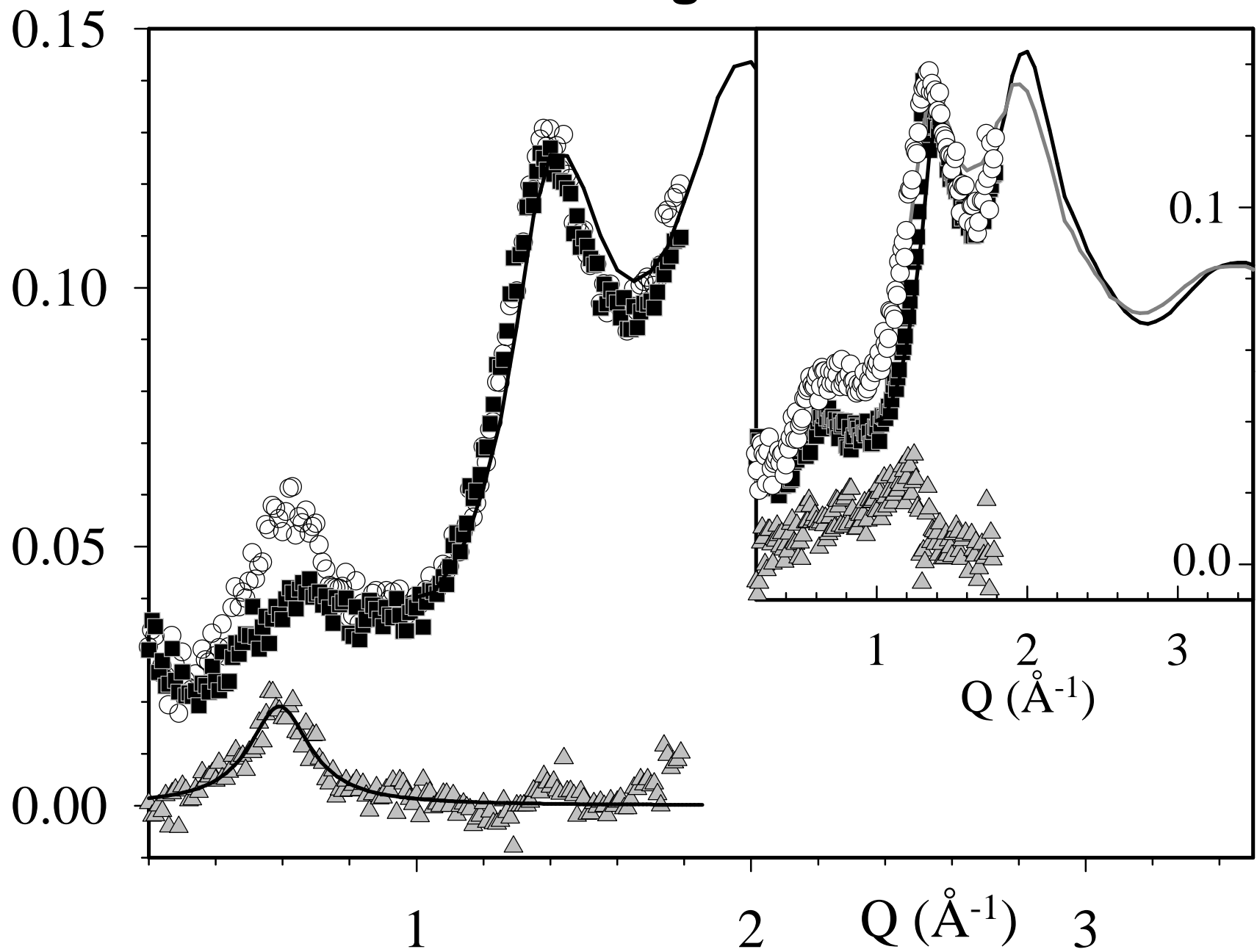
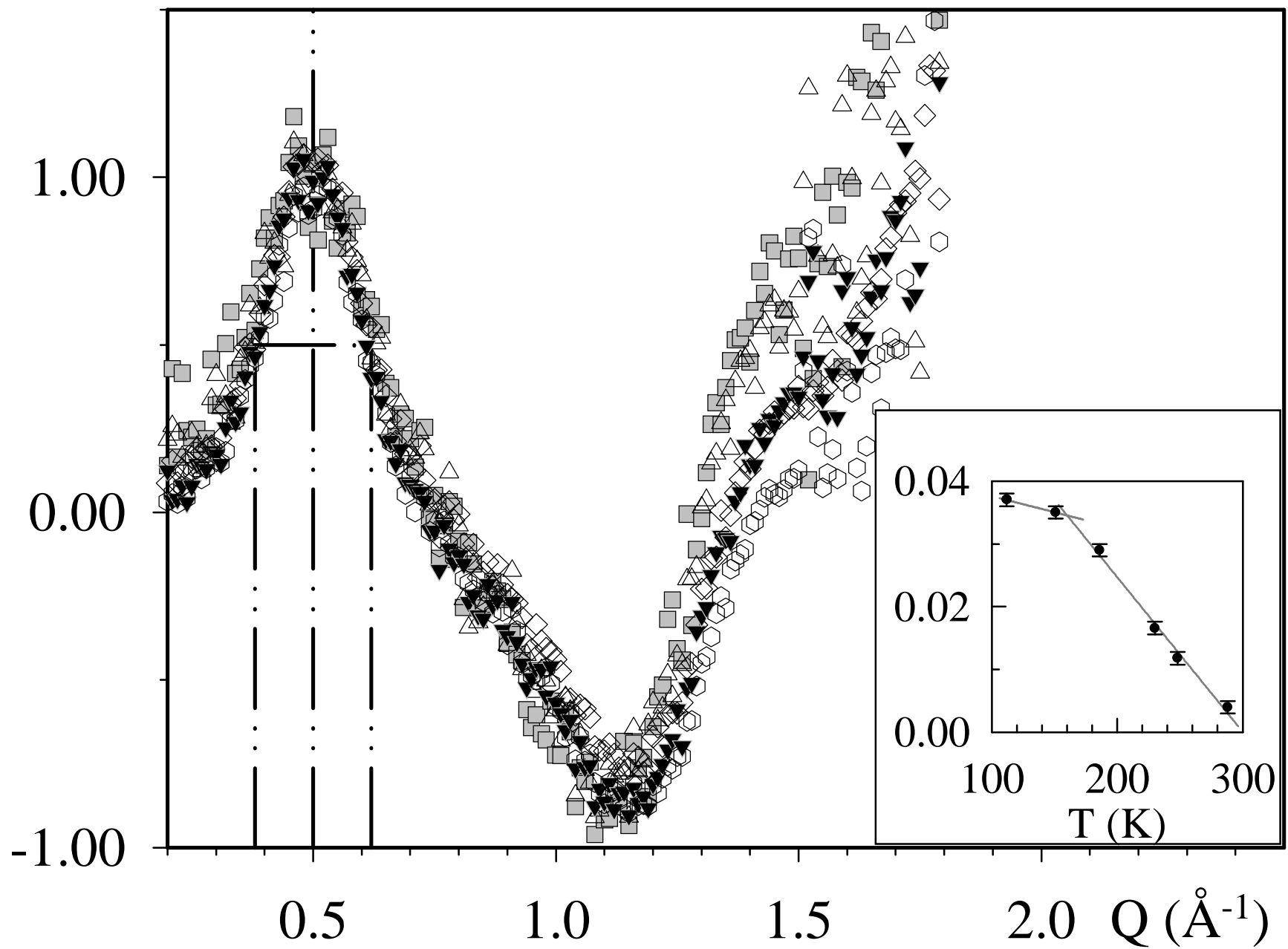
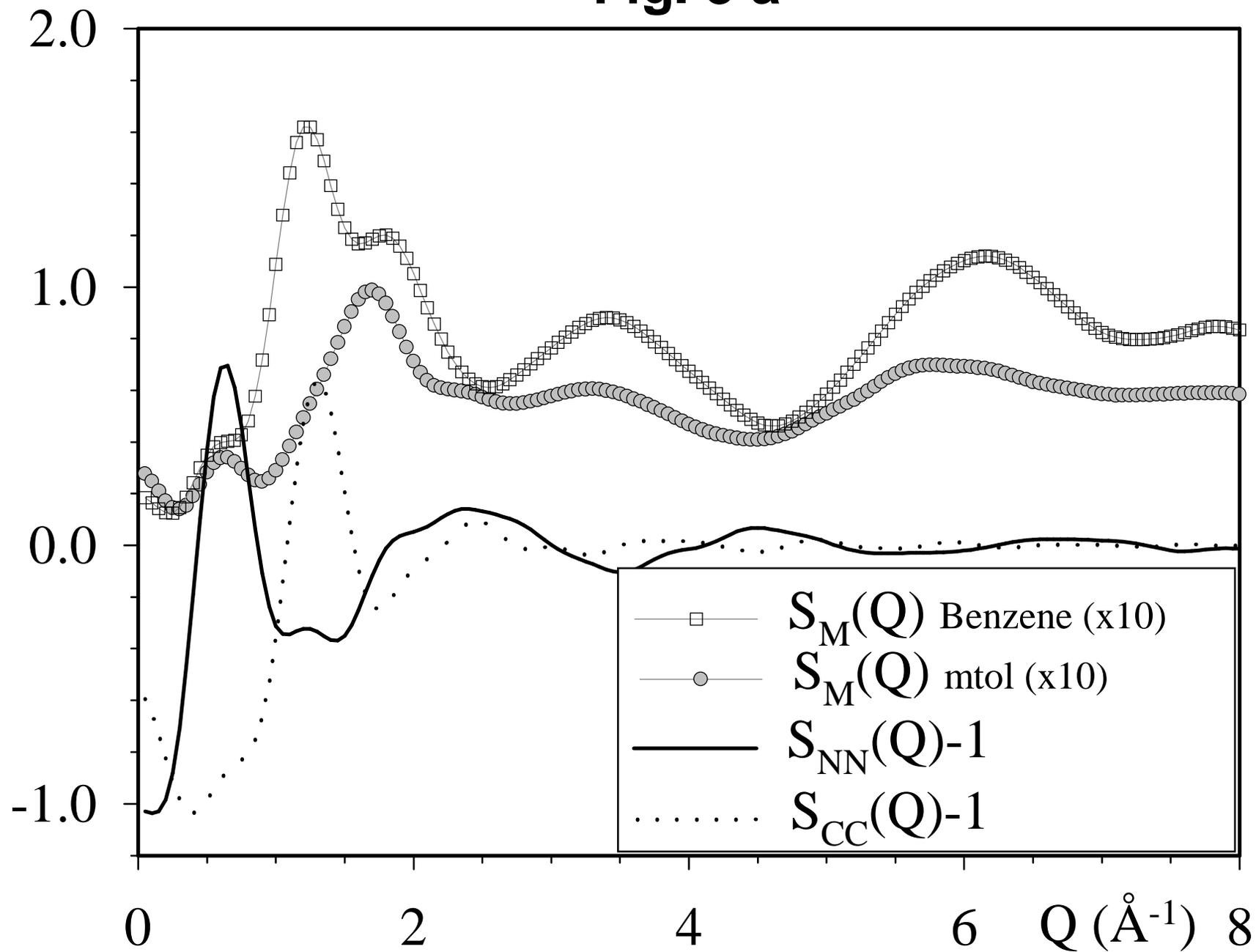
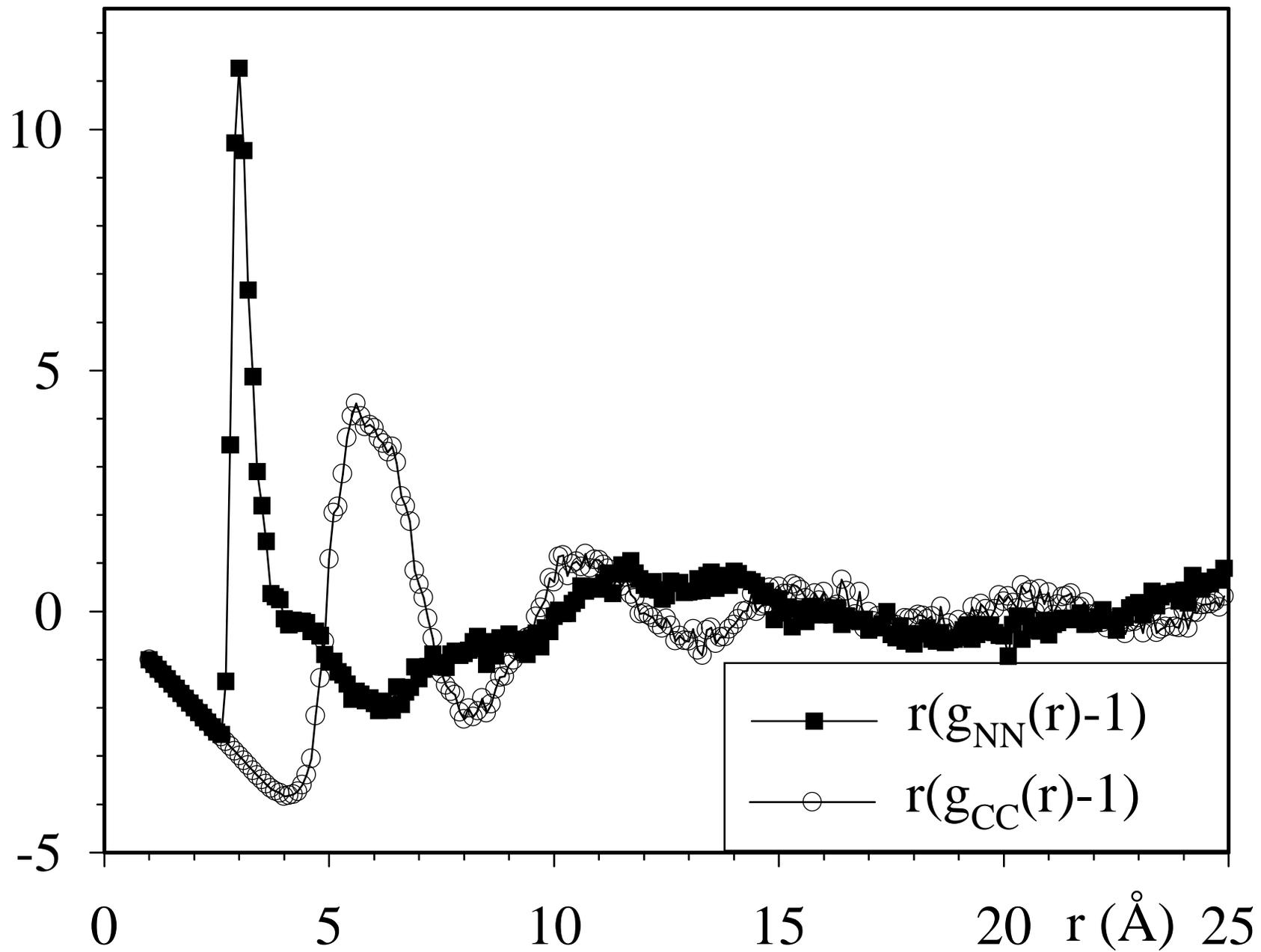
\end{document}